\documentclass[fleqn,usenatbib]{mnras}

\usepackage{newtxtext,newtxmath}

\usepackage[T1]{fontenc}

\DeclareRobustCommand{\VAN}[3]{#2}
\let\VANthebibliography\thebibliography
\def\thebibliography{\DeclareRobustCommand{\VAN}[3]{##3}\VANthebibliography}

\usepackage{graphicx}	
\usepackage{amsmath}	
\usepackage{xspace}
\usepackage{glossaries-extra}
\usepackage{cleveref}
\usepackage[table]{xcolor}
\usepackage{xspace}
\usepackage{booktabs}
\usepackage{placeins}
\usepackage{multirow}
\usepackage{pifont}

\newcommand{\checkme}[1]{{#1}}
\newcommand{\cmark}{\ding{51}}%
\newcommand{\xmark}{\ding{55}}%
\definecolor{lightgrey}{gray}{0.95}

\newcommand{\kld}{D_{\textrm{KL}}}
\newcommand{\jsd}{D_{\textrm{JS}}}

\newcommand{\pocomc}{\texttt{pocomc}\xspace}
\newcommand{\bilby}{\texttt{bilby}\xspace}
\newcommand{\bilbypipe}{\texttt{bilby\_pipe}\xspace}
\newcommand{\dynesty}{\texttt{dynesty}\xspace}

\newcommand{\networksnrtwodet}{\checkme{8.1}\xspace}
\newcommand{\networksnrthreedet}{\checkme{8.7}\xspace}

\newcommand{\efficiencyimprovementtwodet}{\checkme{$1.92 \pm 0.51$}\xspace}
\newcommand{\efficiencyimprovementthreedet}{\checkme{$2.09 \pm 0.56$}\xspace}
\newcommand{\timeimprovementtwodet}{\checkme{$2.63 \pm 0.93$}\xspace}
\newcommand{\timeimprovementthreedet}{\checkme{$2.84\pm 1.04$}\xspace}
\newcommand{\efficiencymean}{\checkme{2}\xspace}
\newcommand{\timemean}{\checkme{2.74}\xspace}

\setabbreviationstyle[acronym]{long-short}

\newacronym{smc}{SMC}{sequential Monte Carlo}
\newacronym{mcmc}{MCMC}{Markov Chain Monte Carlo}
\newacronym{bbh}{BBH}{binary black hole}
\newacronym{bns}{BNS}{binary neutron star}
\newacronym{jsd}{JSD}{Jensen–Shannon divergence}
\newacronym{kld}{KLD}{Kullback-Leibler Divergence}
\newacronym{cbc}{CBC}{compact binary coalenscence}
\newacronym{psd}{PSD}{power spectral density}
\newacronym{lvk}{LVK}{LIGO-Virgo-KAGRA Collaboration}
\newacronym{snr}{SNR}{signal-to-noise ratio}
\newacronym{gwtc}{GWTC}{Gravitational Wave Trasient Catalog}
\newacronym{roq}{ROQ}{reduced order quadrature}
\newacronym{ps}{PS}{persistent sampling}
\newacronym{ns}{NS}{nested sampling}
\newacronym{pcn}{pCN}{preconditioned Crank-Nicolson}
\newacronym{tpcn}{tpCN}{$t$-preconditioned Crank-Nicolson}
\newacronym{pptest}{P-P test}{probability-probability test}
\newacronym{ess}{ESS}{effective sample size}
\newacronym{iid}{i.i.d}{independent and identically distributed}

\title[Validating Sequential Monte Carlo for Gravitational-Wave Inference]{Validating Sequential Monte Carlo for Gravitational-Wave Inference}

\author[M. J. Williams et al.]{
M. J. Williams,$^{1}$\thanks{E-mail: michael.williams3@port.ac.uk}
M. Karamanis,$^{2,3,4}$
Y. Luo,$^{5}$
U. Seljak$^{2,3,4}$
\\
$^{1}$University of Portsmouth, Portsmouth, PO1 3FX, United Kingdom\\
$^{2}$Berkeley Center for Cosmological Physics, UC Berkeley, CA 94720, USA\\
$^{3}$Department of Physics, University of California, Berkeley, CA 94720, USA\\
$^{4}$Lawrence Berkeley National Laboratory, One Cyclotron Road, Berkeley, CA 94720, USA\\
$^{5}$School of Physics, Peking University, Beijing 100871, People's Republic of China
}

\date{Accepted XXX. Received YYY; in original form ZZZ}

\pubyear{\the\year{}}

\begin{document}
\label{firstpage}
\pagerange{\pageref{firstpage}--\pageref{lastpage}}
\maketitle

\begin{abstract}

\Gls{ns} is the preferred stochastic sampling algorithm for gravitational-wave inference for \glspl{cbc}. It can handle the complex nature of the gravitational-wave likelihood surface and provides an estimate of the Bayesian model evidence. However, there is another class of algorithms that meets the same requirements but has not been used for gravitational-wave analyses: \Gls{smc}, an extension of importance sampling that maps samples from an initial density to a target density via a series of intermediate densities. In this work, we validate a type of \gls{smc} algorithm, called \gls{ps}, for gravitational-wave inference. We consider a range of different scenarios including \glspl{bbh} and \glspl{bns} and real and simulated data and show that \gls{ps} produces results that are consistent with \gls{ns} whilst being, on average, \efficiencymean times more efficient and \timemean times faster. This demonstrates that \gls{ps} is a viable alternative to \gls{ns} that should be considered for future gravitational-wave analyses.

\end{abstract}

\begin{keywords}
gravitational waves -- methods: data analysis --  stars: neutron -- stars: black holes -- transients: black hole mergers -- transients: neutron star mergers
\end{keywords}



\glsresetall

\section{Introduction}

Bayesian inference is a cornerstone of gravitational-wave astronomy. It is used for a wide range of analyses, including characterising the compact binaries observed by the \gls{lvk}~\citep{KAGRA:2021vkt}. However, the nature of gravitational-wave data and these signals means that this inference can be complex, often requiring careful design and tuning of inference algorithms, and it can also be computationally expensive. This latter point is increasingly important as the sensitivity of \gls{lvk} detectors improves~\citep{KAGRA:2013rdx} and the next-generation of detectors are built~\citep{Kalogera:2021bya}.

Different inference algorithms have been used within the gravitational-wave community for performing inference~\citep{Ashton:2018jfp,Veitch:2014wba,Lange:2018pyp}, however, \glsfirst{ns}~\citep{Skilling:2004pqw,Skilling:2006gxv} is currently the de facto algorithm for most analyses~\citep{LIGOScientific:2021usb,KAGRA:2021vkt,Nitz:2021zwj,Wadekar:2023gea}.
\Gls{ns} can handle the complex parameter space that results from the gravitational-wave likelihood, which often includes degeneracies and multimodality. It also provides an estimate of the Bayesian model evidence (marginalized likelihood) that can be used for Bayesian model selection when, for example, performing tests of general relativity~\citep{LIGOScientific:2021sio}.

\Gls{smc} methods (also known as particle filters) are an extension to importance sampling that have a wide range of applications for solving particle filtering problems, including to Bayesian inference~\citep{bernardo2011sequential}.
These methods evolve a set of particles from an initial reference distribution (often the prior distribution) to a target distribution (often the posterior distribution) via a sequence of intermediate distributions.
This sequence is typically constructed either by iteratively introducing data or annealing the target distribution using an effective temperature parameter~\citep{doucet2001introduction}.
This approach has several notable benefits; its adaptive nature makes it well suited to complex, multimodal problems whilst being inherently parallelizable.
It also has connections to \gls{ns}, \cite{2018arXiv180503924S}~showed that \gls{ns} can be reformulated in the \gls{smc} framework.
These properties make such methods an interesting avenue for gravitational-wave inference, however, they have only seen limited use in this area~\citep{karamanis2022accelerating}. 

In this paper, we validate sequential Monte Carlo for analysing gravitational-wave signals from \glspl{cbc} and compare it to results obtained with a \gls{ns} algorithm.
The paper is structured as follows: in \cref{sec:gwinference,sec:smc} we review the fundamentals of gravitational-wave inference for \glspl{cbc} and \gls{smc}; in \cref{sec:pocomc} we describe the specific \gls{smc} algorithm we use, \pocomc, and the changes made to it; in \cref{sec:simlated_data} we validate \pocomc using simulated data and compare results to those obtained using the \gls{ns} algorithm, \dynesty; in \cref{sec:real_data} we apply \pocomc to real \gls{cbc} events from \gls{lvk} observing runs; and in \cref{sec:conclusions} we summarise our findings.

\section{Gravitational-wave inference}\label{sec:gwinference}

For \glspl{cbc}, the observed strain data $\tilde{d}(f)$ is typically modelled in the frequency domain in terms of a frequency-dependent signal $\tilde{h}_{\theta}(f)$ with parameters $\theta$, and frequency-dependent noise $\tilde{n}(f)$:
\begin{equation}
    \tilde{d}(f) = \tilde{h}_{\theta}(f) + \tilde{n}(f).
\end{equation}
In current analyses, the noise is assumed to be stationary, Gaussian and described by a one-sided \gls{psd} $S_n(f)$.
As such, the gravitational-wave likelihood is typically defined as a Gaussian function~\citep{Veitch:2014wba,Thrane:2018qnx}:
\begin{equation}\label{eq:gw_likelihood}
    p(d|\theta, S_n(f)) \propto \exp\left\{ -\frac{1}{2}\langle \tilde{d}(f) - \tilde{h}_{\theta}(f) | \tilde{d} - \tilde{h}_\theta(f) \rangle \right\},
\end{equation}\
where $\langle a | b \rangle$ is the inner product~\citep{Cutler:1994ys,Veitch:2009hd}
\begin{equation}
    \langle a | b \rangle = 2 \int_{0}^{\infty} \frac{\tilde{a}(f)\tilde{b}^{*}(f) + \tilde{a}^{*}(f)\tilde{b}(f)}{S_n(f)} \textrm{d}f.
\end{equation}
The signal in \cref{eq:gw_likelihood} is modelled using a waveform approximant that models the predicted strain data as a function of parameters $\theta$.
This will include parameters that are intrinsic to the source, such as the masses and spins, and those which depend on the observer's location, the extrinsic parameters.
Currently analyses predominantly use quasi-circular waveforms that include spin precession and high-order multipoles~\citep{Varma:2019csw,Pratten:2020ceb,Ramos-Buades:2023ehm}, this results in a 15-dimensional parameter space for \glspl{bbh} and a 17-dimensional parameter space for \glspl{bns} --- the two additional parameters quantify how each neutron star is deformed by tidal effects and are known as the tidal deformabilities $\lambda_{\{1,2\}}$.

In the current generation ground-based detectors, the resulting likelihood surface from \cref{eq:gw_likelihood} can be multimodal and highly correlated in the 15 (17) parameters of interest.
This, along with the computational cost associated with state-of-the-art waveforms (a few to hundreds of milliseconds per evaluation) and scaling with data duration, means inference for \gls{cbc} signals can be computationally expensive, taking anywhere from hours to days.

When analysing real data, it is important to account for the frequency-dependent phase and amplitude error that arises from the imperfect calibration of the interferometric data~\citep{2017PhRvD..96j2001C,Viets:2017yvy}.
Therefore, we model the observed signal $\tilde{h}_{\textrm{obs}}(f)$ in terms of these unknown errors:
\begin{equation}
    \tilde{h}_{\textrm{obs}}(f) = \tilde{h}_{\theta}(f) (1 + \delta A(f))\exp\left[ i \delta \phi(f)\right],
\end{equation}
where $\delta A(f)$ and $\delta \phi(f)$ are the frequency-dependent amplitude and phase errors, respectively.
These errors are expected to be small and smoothly varying~\citep{Viets:2017yvy}, and are typically modelled using cubic spline polynomials with 10 nodes each~\citep{Farr:calibration_errors}.
However, this introduces an additional 20 parameters per detector to the inference problem and any inference algorithm that is to be applied to real data must either include these parameters in the likelihood, or handle them via additional post-processing steps, see e.g. \cite{Payne:2020myg}.
This increased dimensionality can prove particularly challenging for \gls{ns} algorithms that rely on region-based proposal methods~\citep{Williams:2023iss}.


As a result of the challenging nature of gravitational-wave inference, there has been extensive work into accelerating it, covering a wide range of approaches. Some focus on accelerating stochastic sampling and others focus on alternative methods such as simulation based inference~\citep{Dax:2021tsq,Bhardwaj:2023xph} or non-Bayesian approaches~\citep{Fairhurst:2023idl}. In this work, we focus on the former which can be broadly be divided into two categories: those that aim to reduce the cost of computing \cref{eq:gw_likelihood} and those that aim to improve efficiency or scaling of the inference algorithm. This works falls into the latter category.

Within the area of the stochastic sampling algorithms, there has been extensive research into developing algorithms that are more efficient. Some works propose modifications to \gls{ns} or \gls{mcmc} that, for example, incorporate normalizing flows~\citep{Williams:2021qyt,Ashton:2021anp,Wong:2023lgb,Prathaban:2024rmu}. Others focus on developing new algorithms, many of which are based on importance sampling~\citep{Williams:2023ppp,Tiwari:2023mzf,Saleh:2024tgr}. However, not all of these algorithms have been validated on the `full' inference problem: analyses including higher-order multipoles, spin procession and calibration uncertainties. 
In this work, we aim to validate a new sampling algorithm for the full inference problem.

These algorithms are integrated into a range of gravitational-wave inference pipelines that implement the likelihood and associated data processing. Some pipelines use stochastic sampling algorithms --- such as \bilby~\citep{Ashton:2018jfp,Romero-Shaw:2020owr}, \texttt{LALInference}~\citep{Veitch:2014wba}, \texttt{PyCBC Inference}~\citep{Biwer:2018osg} and \texttt{cogwheel}~\citep{Roulet:2022kot,Islam:2022afg,Roulet:2024hwz} --- whilst others use alternative algorithms including \texttt{RIFT}~\citep{Lange:2018pyp} and \texttt{DINGO}~\citep{Dax:2021tsq}.
The most recent \gls{gwtc}, GWTC-3~\citep{KAGRA:2021vkt} used a combination of \bilby~\citep{Ashton:2018jfp}(with the \dynesty sampler~\citep{Speagle:2019ivv}) and \texttt{RIFT}~\citep{Lange:2018pyp}.

\section{Sequential Monte Carlo and Persistent Sampling}\label{sec:smc}

\begin{figure}
    \centering
    \includegraphics{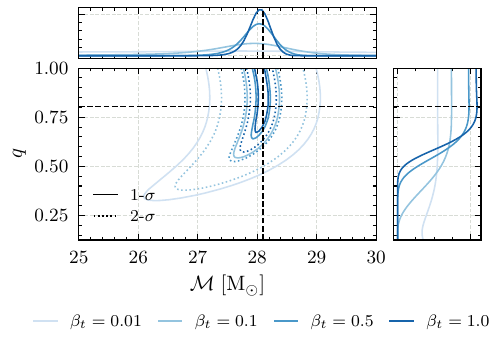}
    \caption{Annealed one- and two- dimensional posterior distributions (see \cref{eq:intermediate_densities}) for chirp mass ($\mathcal{M}$) and mass ratio ($q$) for a GW150914-like injection in zero-noise at four different inverse temperatures ($\beta_t$). The injection was simulated in a two-detector network using \texttt{IMRPhenomPv2}. The same waveform was used when computing the likelihood and all other parameters were fixed to the injected values except for the phase at coalescence, which as analytically marginalized~\citep{Thrane:2018qnx}. The dashed vertical and horizontal lines indicate the injection parameters.}
    \label{fig:annealing_example}
\end{figure}

\Glsfirst{smc} samplers iteratively transform an initial set of samples (called particles), typically drawn from the prior distribution, into a set of samples representing the target posterior distribution through a sequence of intermediate steps. A common and effective strategy to define these intermediate distributions is temperature annealing. One can conceptualize the likelihood function $p(d|\theta)$ as being gradually introduced. This is controlled by an inverse temperature parameter, $\beta_t$, that increases from $\beta_1 = 0$ (where the distribution is simply the prior) to $\beta_T = 1$ (where the distribution is the full posterior). At each step $t$, the intermediate distribution $p_t(\theta)$ is defined as:
\begin{equation}\label{eq:intermediate_densities}
    p_t(\theta) = \frac{p(d|\theta)^{\beta_t}p(\theta)}{Z_{t}},
\end{equation}
where $p(\theta)$ is the prior and $Z_t = \int p(d|\theta)^{\beta_t}p(\theta) d\theta$ is the normalization constant (or evidence) at that temperature.
Thus, $p_1(\theta)$ is the prior $p(\theta)$ (since $p(d|\theta)^0=1$), and $p_T(\theta)$ is the target posterior $p(\theta|d) \propto p(d|\theta)p(\theta)$. The final normalization constant $Z_T$ is the standard Bayesian evidence. This gradual "annealing" process is particularly advantageous as it facilitates the algorithm's navigation of complex likelihood surfaces. Such surfaces may feature multiple, well-separated modes (peaks) or narrow, convoluted ridges, which can be difficult for algorithms attempting to sample the posterior directly, as they risk becoming trapped in a single local optimum, thereby failing to map the full posterior landscape. Temperature annealing specifically addresses the challenge of multimodality because, at low $\beta_t$ values (corresponding to high effective temperatures), the posterior landscape is significantly broadened and smoothed. In this flattened state, the probabilistic barriers that might ordinarily isolate different modes are substantially reduced. This allows particles to traverse the parameter space more freely, making it easier for the sampler to discover and populate disparate regions of high probability that might otherwise be inaccessible. As $\beta_t$ subsequently increases (i.e., the effective temperature decreases), these initially widespread particles, now hopefully distributed across various discovered modes, can then settle into and refine the characterization of each distinct mode as the features of the true posterior sharpen. As illustrated in \cref{fig:annealing_example}, at low $\beta_t$, the annealed posterior is broad, which simplifies initial exploration; as $\beta_t$ increases, the distribution progressively concentrates, converging towards the true posterior and its potentially complex modal structure.

\begin{figure}
    \centering
    \includegraphics[scale=0.079]{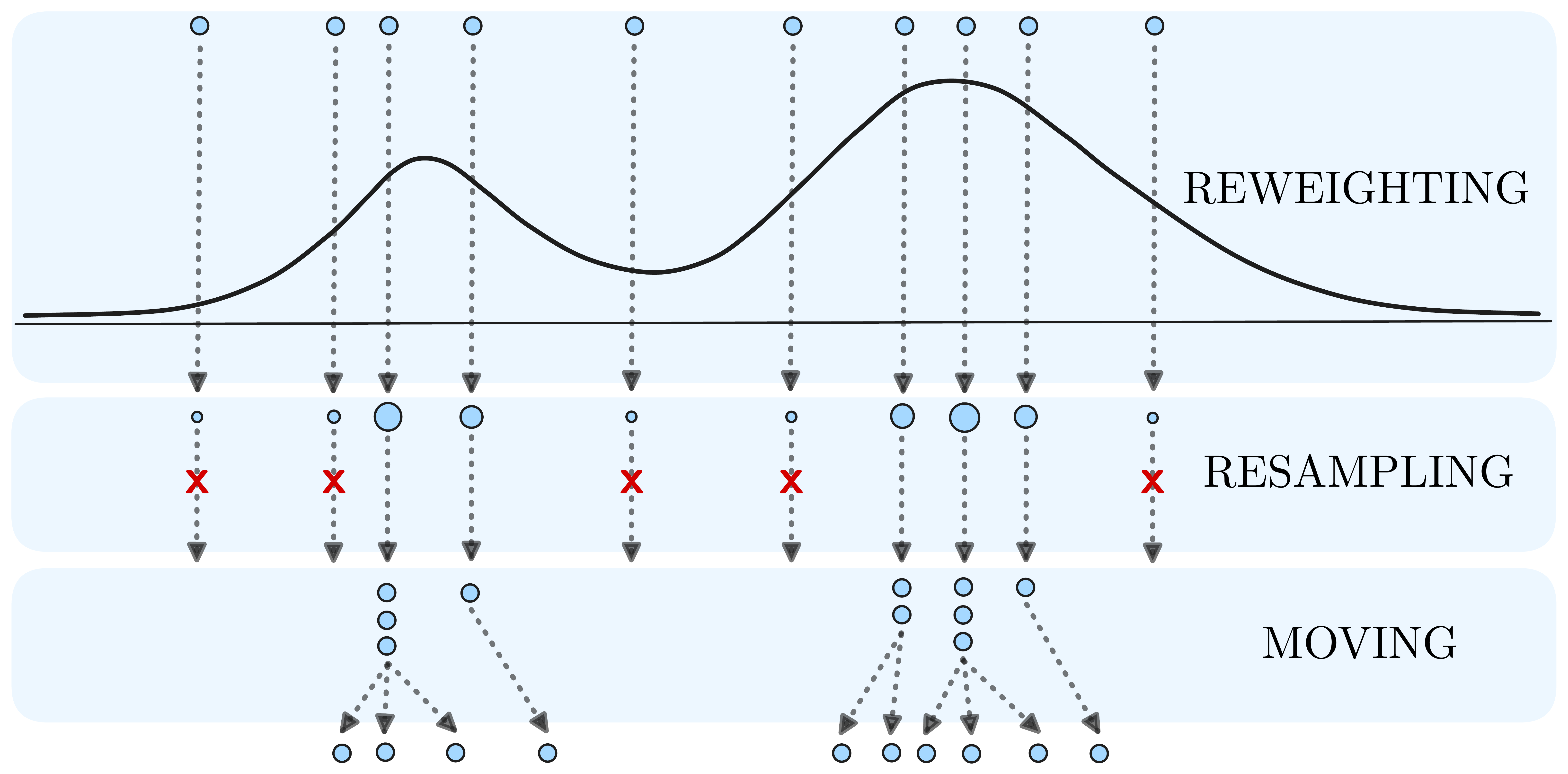}
    \caption{Single iteration $t$ of the \gls{smc} algorithm demonstrating its three key operations, (i) reweighting, (ii) resampling, and (iii) moving. Particle positions are shown in 1D with particle weights represented by their relative size. Low-weight particles are discarded and high-weight ones are multiplied during resampling. The moving step diversifies the resampled population, yielding equally weighted and distinct particles.}
    \label{fig:smc_iteration}
\end{figure}

To transition the $N$ particles, $\{\theta_{t-1}^i\}_{i=1}^{N}$, from approximating distribution $p_{t-1}(\theta)$ to approximating the next distribution $p_t(\theta)$, standard \gls{smc} algorithms employ three key operations (illustrated in \cref{fig:smc_iteration}) at each iteration $t$:
\begin{enumerate}
    \item{
        \textbf{Reweighting:} Each particle $\theta_{t-1}^i$, which was a sample from $p_{t-1}(\theta)$, is assigned an importance weight. This weight quantifies how well the particle fits the new target distribution $p_t(\theta)$ compared to how it fit the old one $p_{t-1}(\theta)$. The unnormalized weight for particle $i$ is given by
        \begin{equation}
            w_t^i = p(d|\theta_{t-1}^i)^{\beta_t - \beta_{t-1}},
        \end{equation}
        (derived from $p_t(\theta_{t-1}^i)/p_{t-1}(\theta_{t-1}^i)$). These weights guide the particle population towards regions of higher probability under the new, more concentrated distribution $p_t(\theta)$. The normalized weights $W_t^i$ are proportional to $w_t^i$. The ratio of the normalization constants, $Z_t/Z_{t-1}$, is estimated as the average of these unnormalized weights: 
        \begin{equation}
            \frac{\hat{Z}_t}{\hat{Z}_{t-1}} = \frac{1}{N}\sum_{j=1}^{N}w_t^j.
        \end{equation}
        Given that $Z_1=1$ (if the prior is normalized), this allows for an iterative estimation of the final evidence $Z_T = Z_1 \times (\hat{Z}_2/\hat{Z}_1) \times \dots \times (\hat{Z}_T/\hat{Z}_{T-1})$. The temperature $\beta_t$ is often determined adaptively at each step by, for example, requiring that the \gls{ess}~\citep{kish_ess} of the weighted particles remains above a certain threshold, ensuring sufficient diversity of weights.
    }
    \item{\textbf{Resampling:} After reweighting, some particles will have very low weights, indicating they are in regions of lesser importance for $p_t(\theta)$. To focus computational effort efficiently, a resampling step is performed: $N$ new particles are drawn (with replacement) from the current set $\{\theta_{t-1}^i\}$ with probabilities proportional to their normalized weights $\{W_t^i\}$. This step tends to discard particles with low weights and replicate those with high weights, thereby concentrating the particle population in the more relevant regions of the parameter space for $p_t(\theta)$.
    }
    \item{\textbf{Moving:} The resampling step can lead to multiple identical copies of high-weight particles. To diversify the particle population and allow them to explore the current distribution $p_t(\theta)$ more thoroughly, each newly resampled particle $\tilde{\theta}_t^i$ is then perturbed or "moved". This is typically done by applying an \gls{mcmc} kernel, $\mathcal{K}_t$, for one or more steps, where $\mathcal{K}_t$ is chosen to have $p_t(\theta)$ as its stationary distribution. This step, resulting in the updated particles $\theta_t^i \leftarrow \mathcal{K}_t(\tilde{\theta}_t^i)$, helps the particles explore the local landscape of $p_t(\theta)$ and resolve degeneracies. The choice of \gls{mcmc} kernel (e.g., random-walk Metropolis~\citep{Metropolis:1953am}, Hamiltonian Monte Carlo~\citep{Neal:2011}) can be adapted based on the properties of the current particle set, such as their empirical covariance, which is an advantage of the \gls{smc} framework.
    }
\end{enumerate}
Standard \gls{smc} addresses many shortcomings of simpler importance sampling; however, it still has limitations, particularly related to the fixed number of particles, which can lead to high variance in estimates and challenges like particle impoverishment (loss of diversity after resampling). As such, various modifications to this standard algorithm have been proposed. In this work, we consider one specific approach called \glsxtrlong{ps}. For an in-depth introduction to \gls{smc}, see \cite{doucet2001introduction}.

\Glsfirst{ps}, introduced by \cite{karamanis2025}, is an extension of \gls{smc} specifically designed to address significant limitations of standard \gls{smc} such as particle impoverishment and mode collapse, especially when computational resources are constrained. While \gls{ps} also navigates a sequence of annealed distributions $p_t(\theta)$, its principal innovation is its particle management strategy. Instead of discarding particles from previous iterations ($t'<t$) after each resampling phase, \gls{ps} systematically retains and reuses particles from all prior iterations. This process constructs a growing, weighted ensemble of particles that are all utilized to represent the target distribution at the current iteration $t$.

The underlying mechanism involves treating the accumulated particles from all previous iterations $s=1, \dots, t-1$ (a total of $(t-1) \times N$ particles if $N$ particles are generated per iteration) as approximate samples from a mixture distribution 
\begin{equation}
    \tilde{p}_t(\theta) = \frac{1}{t-1}\sum_{s=1}^{t-1}p_s(\theta)\,.
\end{equation}
Reweighting is then performed using principles from multiple importance sampling, assigning new weights to all these $(t-1) \times N$ historical particles so that they collectively target the current distribution $p_t(\theta)$. From this much larger, combined pool, $N$ new active particles are resampled, which are then diversified through the standard moving step.

The "persistent" methodology, as implemented in \Gls{ps}, offers substantial advantages, both as an enhancement to standard \gls{smc} techniques and when compared to other classes of algorithms such as \gls{ns}.

\Gls{ps} significantly enhances standard \gls{smc} methodologies by systematically leveraging a growing, persistent pool of all previously generated particles. This core strategy leads to several key improvements~\citep{karamanis2025}: \Gls{ps} achieves higher accuracy efficiently, as reweighting historical particles utilizes cached likelihood values, thereby avoiding new computationally expensive likelihood evaluations for these samples. The resulting larger and more diverse particle ensemble allows for the construction of more robust posterior approximations, directly mitigating common \gls{smc} issues such as particle impoverishment and mode collapse. Consequently, estimates of marginal likelihood and posterior expectations typically exhibit significantly lower variance. Furthermore, the process of resampling active particles from this extensive persistent pool yields new particles that are naturally more distinct and less correlated, which alleviates the diversification burden on the \gls{mcmc} moving step and facilitates more stable and efficient adaptation of its transition kernels.

When contrasted with other methodologies like \gls{ns}, \gls{ps} with annealing also presents several distinct advantages:
\begin{itemize}
    \item {
        \textbf{Highly Efficient Parallelization:} The \gls{mcmc} moving step, central to \gls{ps} (inherited from \gls{smc}), is embarrassingly parallel. This allows for near-linear performance scaling with the number of processing units, offering a significant advantage over methods like NS which often have more constrained parallelization capabilities.
    }
    \item {
        \textbf{Favourable High-Dimensional Scaling:} \Gls{ps} with temperature annealing reportedly exhibits computational cost scaling with parameter space dimensionality ($D$) as approximately $\mathcal{O}(D^{1/2})$~\citep{chopin2020introduction}. This offers a potential performance advantage in high-dimensional problems compared to algorithms with less favourable scaling (e.g., the typical $\mathcal{O}(D)$ scaling for NS), particularly when the posterior is significantly more concentrated than the prior.
    }
    \item {
        \textbf{Enhanced Exploration of Complex Posteriors via Annealing:} Temperature annealing systematically enables \gls{ps} to effectively navigate complex, multimodal likelihoods. It allows particles to initially explore a smoothed landscape (at low $\beta_{t}$), aiding the discovery of disparate modal regions before the posterior sharpens as $\beta_{t}\to 1$. While NS also addresses multimodality, the annealing mechanism in \gls{ps} provides a structured, global-to-local exploration strategy.
    }
    \item {
        \textbf{Natural Termination and Online Learning Adaptability:} Unlike NS, the annealing process in \gls{ps} has a clear termination point when $\beta_{t}$ reaches its target (typically 1). Furthermore, the underlying \gls{smc} framework inherent to \gls{ps} directly supports online learning via data tempering (sequential data introduction), making it suitable for streaming data applications, a feature less readily available in standard NS implementations.
    }
\end{itemize}
These characteristics establish \gls{ps} as a robust, scalable, and efficient alternative for addressing complex Bayesian inference tasks, demonstrating superior performance compared to standard \gls{smc} and related variants, particularly under constrained computational budgets. This makes it an interesting candidate for gravitational-wave inference~\citep{karamanis2022accelerating}. For a comprehensive description of the \gls{ps} algorithm, its theoretical underpinnings, and empirical performance, readers are referred to \cite{karamanis2025}.

\section{\pocomc}\label{sec:pocomc}

In this work, we use a specific implementation of \gls{ps}, called \pocomc~\citep{karamanis2022accelerating,karamanis2022pocomc}, this includes the version of the algorithm described in \cite{karamanis2025}.
\pocomc includes several features that make it well suited to complex inference problems, such as gravitational-wave inference for \glspl{cbc}. These include, but are not limited to:
\begin{itemize}
    \item {
        \textbf{Preconditioning with normalizing flows:} \pocomc leverages normalizing flows~\citep{2019arXiv190809257K,2019arXiv191202762P} to apply a change-of-variables transformation (preconditioning) to the moving step. The preconditioning reduces the correlation between parameters, which improves the efficiency of the \gls{mcmc} sampling and reduces the reliance on an appropriate choice of proposal distribution. This has parallels to how normalizing flows have been used in other sampling algorithms for gravitational-wave inference, such as \texttt{nessai}~\citep{Williams:2021qyt,Williams:2023ppp} and \texttt{flowMC}~\citep{Wong:2022xvh,Gabrie:2021tlu}, which have been shown to improved sampling efficiency for gravitational-wave problems compared to standard algorithms~\cite{Williams:2021qyt,Williams:2023ppp,Wong:2023lgb}.
    }
    \item {
        \textbf{$t$-preconditioned Crank-Nicolson \gls{mcmc}}: \pocomc implements the \gls{tpcn} \gls{mcmc} algorithm~\citep{2024InvPr..40l5023G}, an \gls{mcmc} sampling scheme that is well suited to high-dimensional problems where the target distribution is non-Gaussian and that out-performs other, simpler \gls{mcmc} methods. \cite{karamanis2025} applied \pocomc with this \gls{mcmc} algorithm to range of challenging toy problems and demonstrated its robustness. This is particularly important since previous studies have shown that simpler \gls{mcmc} algorithms are insufficient for gravitational-wave inference problems~\citep{Kulkarni:2020htf} and more complex algorithms are often required~\citep{Veitch:2014wba,Ashton:2021anp}.
    }
    \item {
        \textbf{Parallelization:} as mentioned previously, the moving step in \gls{smc} is embarrassingly parallel --- the speed-up scales approximately linearly with number of processes --- and \pocomc supports such parallelization. Furthermore, typical applications can use upwards of 1000 particles meaning that the algorithm can easily scale to hundreds of parallel processes. \Gls{ns}, in contrast, exhibits are diminishing returns as the number of parallel process increases~\citep{2015MNRAS.453.4384H}. In this work, we make use of the support for external parallelization and use the pool provided by \bilby, though \pocomc supports various forms of parallelization~\cite{pocomc_docs}.
    }
    \item {
        \textbf{User-specified number of posterior samples:} \pocomc allows the user to specify the desired number of effective posterior samples independently of the number of particles for the \gls{smc} algorithm. This is achieved by drawing additional samples once the main \gls{smc} algorithm is complete and avoids the need to increase the number of particles. When using \pocomc this is specified via the \texttt{n\_total} argument, see the documentation for more details~\citep{pocomc_docs}.
    }
\end{itemize}

We also make a series of improvements to \pocomc which are designed to handle certain characteristics of the gravitational-wave parameter space, however, they are not problem-specific. 
The changes are:
\begin{itemize}
    \item {
        \textbf{Sliced Iterative Normalizing Flows}:
        instead of the standard Masked Autoregressive Flows, we use Sliced Iterative Normalizing Flows~\citep{dai2021sliced}. We find that these flows are better at handling multimodal distributions than the Masked Autoregressive Flows~\citep{2017arXiv170507057P} and Real Non-Volume Preserving flows~\citep{2016arXiv160508803D} previously available in \pocomc. In the version used in this work, the flows are implemented in \texttt{sinflow}~\citep{sinflow} and are used by default when running \pocomc.
    }
    \item {
        \textbf{Boundary conditions}:
        since a number of the gravitational-wave parameters have periodic boundary conditions, we implement periodic boundary conditions in \pocomc. Specifically, we map the angles to $[0, 2\pi)$ and apply a rotation such the mean is $\pi$. We find this reduces undersampling at the prior bounds for periodic parameters that can otherwise occur.
    }
\end{itemize}

\begin{figure}
    \centering
    \includegraphics{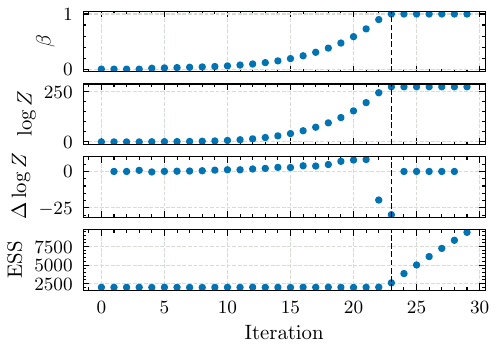}
    \caption{Evolution of the $\beta$ parameter, log-evidence $\log Z$ and \gls{ess} of the particles when analysing GW150914\_095045 using \pocomc, this analysis includes calibration uncertainties. The $\beta$ parameter is adaptively chosen at each iteration and slowly increases until $\beta=1$, during this time, the \gls{ess} of particles is kept constant. Once $\beta=1$ additional samples are drawn to reach the \gls{ess} specified by the user, 10,000 in this case. The results are discussed in detail in \cref{sec:real_data}.}
    \label{fig:smc_history}
\end{figure}

\Cref{fig:smc_history} shows an example of how the inverse temperature parameter, $\beta$, evolves when applied to a \gls{cbc} analysis. It initially evolves slowly whilst the sampler has yet to explore the bulk of the posterior, and then increases more rapidly at later iterations. The log-evidence follows a similar trend, rapidly increasing as $\beta \to 1$ but then decreasing over the final iterations as it converges to the final value.

In order to perform gravitational-wave analyses, we implement an interface between \pocomc and the Bayesian inference library, \bilby~\citep{Ashton:2018jfp,Romero-Shaw:2020owr,colm_talbot_2025_15059020} in the \texttt{pocomc-bilby} package~\citep{pocomc_bilby}. This leverages the sampler-plugin interface in \bilby to enable the user to run \pocomc as they would any other sampler in \bilby via the \texttt{run\_sampler} interface or via \bilbypipe. The source code is available via PyPI and at \url{https://github.com/mj-will/pocomc-bilby} which also includes examples of how to run \pocomc via \bilby.

\section{Validating \pocomc for gravitational-wave inference}\label{sec:simlated_data}

We validate \pocomc using simulated data.
All analyses are run using \bilby and \bilbypipe~\citep{Ashton:2018jfp,Romero-Shaw:2020owr,colm_talbot_2025_15059020} and \texttt{pocomc-bilby} plugin described previously. Code to reproduce all of the analyses and the results are available in the accompanying data release~\citep{code_release,data_release}.

We simulate signals from compact binary coalescences in LIGO Livingston (L), LIGO Hanford  (H) and Virgo (V) at O3 sensitivity~\cite{LIGOScientific:2014pky,VIRGO:2014yos,KAGRA:2013rdx} and consider both three- (HLV) and two-detector (HL) networks.
We compare our results with those obtained with \bilby with the \dynesty~\citep{Speagle:2019ivv} nested sampler; we specifically use the modified version of \dynesty included in \bilby\footnote{For details, see \url{https://bilby-dev.github.io/bilby/dynesty-guide.html}}.

\subsection{Binary black hole analyses}\label{sec:bbh}

\begin{figure}
    \centering
    \includegraphics{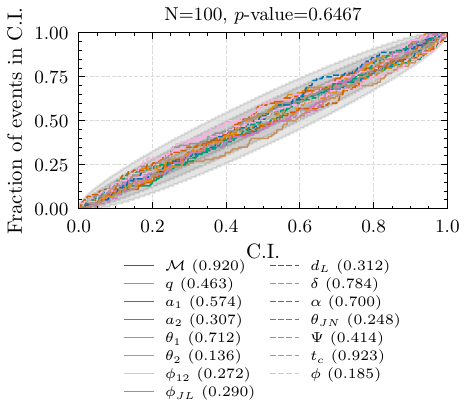}
    \caption{Probability-probability plot obtained using \pocomc to analyse 100 simulated \gls{bbh} signals in two-detector network consisting of LIGO Hanford and LIGO Livingston. The shaded regions denote the 1-, 2- and 3-$\sigma$ bounds respectively. Per-parameter $p$-values are quoted, as well as the combined $p$-value computed using Fisher's method~\citep{fisher1970statistical}.}
    \label{fig:pptest}
\end{figure}

For \gls{bbh} simulations, we employ the waveform approximant \texttt{IMRPhenomXPHM}~\citep{Pratten:2020ceb} and simulate 100 signals in eight seconds of coloured Gaussian noise.
We randomly draw the parameters from the priors described in \cref{app:priors}, which include detector-frame chirp masses in $[25, 35]\;\textrm{M}_{\odot}$.
The resulting signals have a median network \gls{snr} of \networksnrtwodet and \networksnrthreedet in the two- and three-detector networks, respectively.
We run the analyses with distance marginalization enabled and use 4 cores per analysis.
We also use the detector-based sky parameterization (azimuth $\epsilon$ and zenith $\kappa$) and define the reference time at the detector with the highest \gls{snr} ($t_\textrm{IFO}$)~\citep{Romero-Shaw:2020owr}.

We tune the sampler settings for \pocomc for the two-detector analyses, since we observe that these are more challenging than the three-detector analyses due to the correlations that arise in extrinsic parameters. These settings are detailed in \cref{tab:settings_pocomc_bbh}. Unlike \gls{ns}, \pocomc allows the user to specify the desired number of effective posterior samples, we set this to 10,000 per event, however, for consistency with other samplers in \bilby, we perform rejection sampling to produce \gls{iid} samples meaning that in practice we obtain fewer than 10,000 samples.

We perform probability-probability test for both network configurations and present the probability-probability plots for the two-detector case in \cref{fig:pptest}. 
Results for the three-detector case are included in \cref{app:additional_results}.
For each parameter, these plots show the fraction of events for which the injected value lies above the $n\%$ confidence interval. If the results are unbiased, then for $n\%$ confidence interval, $n\%$ of the events should lie above it (plus some uncertainty) and the line for a given parameter should lie near the diagonal.
In both cases, the results obtained with \pocomc are consistent with being unbiased; the results for all 15 parameters lie within the 3-$\sigma$ uncertainties and the combined $p$-values computed using Fishers's method~\citep{fisher1970statistical} are greater than 0.05.

\begin{figure}
    \centering
    \includegraphics{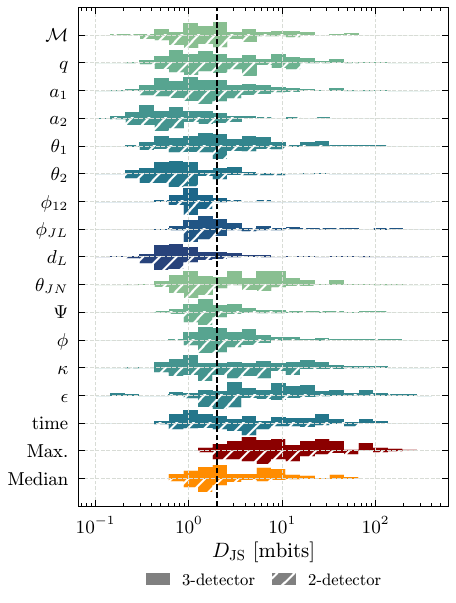}
    \caption{Per-parameter \gls{jsd} between the posterior samples obtained with \pocomc and \dynesty for 100 simulated binary black hole signals when analysed in two- and three-detector networks. The upper and lower histograms show the three- and two-detector results respectively, the counts have been rescaled such the bin with the most counts has the same height between different parameters. The vertical dashed line shows the threshold proposed in \protect\cite{Ashton:2021anp}. `Max.' and `Median' denote the maximum and median \gls{jsd} over all parameters per analysis, respectively. $t_\textrm{IFO}$ denotes the time of coalescence as measured in the detectors with highest \gls{snr}.}
    \label{fig:bbh:jsd}
\end{figure}

To further validate the results, we analyse the same injections using \dynesty, as implemented in \bilby with the settings detailed in \cref{tab:settings_dynesty_bbh}, and compare the posteriors to those obtained with \pocomc.
Probability-probability plots for the two- and three-detectors are shown in \cref{app:additional_results}.
With the settings used, \dynesty passes both tests but the individual $p$-values for some of the parameters are lower than for \pocomc, most notably for the declination $\delta$ and time of coalescence $t_\textrm{c}$.
We attribute this to the signals being relatively low \gls{snr}, including some that would likely not be detected by a matched-filter based search, which the settings are not tuned for. Furthermore, in \gls{lvk} analyses \dynesty would be run multiple times and the result combined together, thus improving the robustness of the results.

We quantify the difference between the inferred posterior samples by following previous works and computing the \gls{jsd} between them. This measures the similarity between two probability distributions as a value between 0 and 1--- larger values correspond to larger differences. The \gls{jsd} is defined as
\begin{equation}
    \jsd[p||q] = \frac{\kld[p||m] + \kld[q||m]}{2},
\end{equation}
where $\kld(p||m)$ the \gls{kld} and $m=0.5(p + q)$ and the units will depend on the base on the logarithm used to compute $\kld$. In this work, we use base 2 which corresponds to bits. Ideally, one would compute the $n$--dimensional \gls{jsd}, however, this is challenging in practice and we instead follow previous studies and compute the \gls{jsd} for the marginal posterior distributions. In practice, this requires fitting a kernel density estimator to each set of samples, computing the probability on a regular grid and then using these to compute the \gls{jsd}.

We report the \gls{jsd} between the marginal posterior distributions obtained with \pocomc and \dynesty in \cref{fig:bbh:jsd}.
This shows that there are larger differences between the samplers for the extrinsic parameters with sky parameters, azimuth $\epsilon$ and zenith $\kappa$, and time of coalescence being the worse overall.
\cite{Ashton:2021anp} propose threshold of $10 / n_\text{eff}\;\textrm{mbit}$ to determine if the samples are consistent or not, we compute the \gls{jsd} using a random subset of 5,000 samples, so the corresponding threshold is $2\;\textrm{mbit}$.
However, we note that the \gls{jsd} values can change significantly depending on the settings used in its calculation, such as the number of bins and the type of kernel density estimator used\footnote{Our implementation uses different kernel density estimators for periodic, bounded and unbounded parameters and we find this significantly impacts the \gls{jsd} values compared to implementations do not account for boundaries.}.  This means that while \gls{jsd} is a useful comparative measure, comparing its absolute value between different works may not be informative, which brings into question the validity of the $10 / n_\text{eff}\;\textrm{mbit}$ threshold.
The $2\;\textrm{mbit}$ threshold implies that many of the posterior distributions are inconsistent between the two samplers; however, the exact nature of the inconsistencies can vary greatly.
We examine the posterior distributions and identify the most common differences:
\begin{itemize}
    \item {Different weights between modes in multimodal posterior distributions, this is most prevalent in sky and time parameters which at these \glspl{snr} are often highly multimodal. \pocomc often assigns larger weights to secondary modes in e.g. the sky localization than \dynesty;}
    \item {Different posterior probability at the prior bounds for intrinsic parameters that significantly rail against the prior bounds. This is a by-product of requiring the same priors for all injections when performing a probability-probability test and of the low \glspl{snr} being considered. In more realistic analyses, the prior bounds would be updated to avoid this;}
    \item {Narrow posteriors where the \gls{jsd} calculation appears to breakdown.}
\end{itemize}
We therefore concluded the differences do not necessarily suggest that either sampler is incorrect but that analyses at low \glspl{snr} can be challenging. In practice, such analyses are uncommon since many of these injections would not be recovered by searches and, as mentioned previously, would combine multiple analyses to mitigate differences between different sampling realisations.

\begin{figure}
    \centering
    \includegraphics{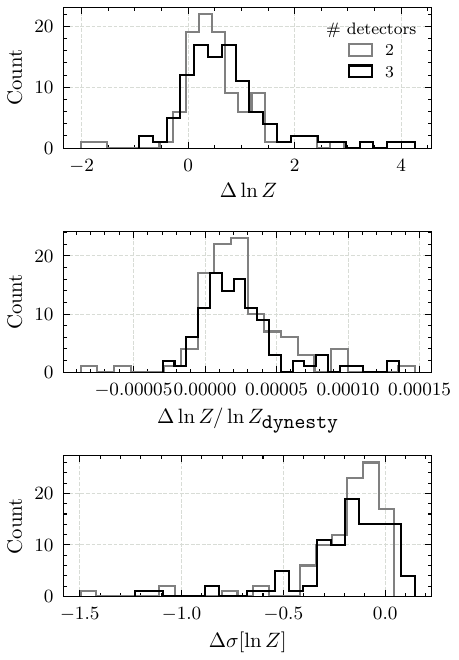}
    \caption{Comparison of the log-evidences estimated using \dynesty and \pocomc in 2- and 3-detector network configurations. The log-evidence estimates for \pocomc were computed using 10,000 samples. All differences are defined as \dynesty--\pocomc. \textbf{Top:} absolute difference between estimated log-evidence. \textbf{Middle:} relative different difference in the log-evidence estimates. \textbf{Bottom:} absolute difference in the estimated uncertainty on the log-evidence.}
    \label{fig:bbh_log_evidence}
\end{figure}

We can also compare the log-evidence estimates returned by each sampler. If the results are consistent, then any differences should be smaller than the associated uncertainty. However, since we observe differences in in the inferred posterior distributions, we should expect to also see differences in the log-evidences.
\Cref{fig:bbh_log_evidence} shows the distribution of absolute and relative differences in estimated log-evidences between \dynesty and \pocomc. The distribution is skewed towards positive differences, suggesting that \pocomc tends to under-estimate the log-evidence compared to \dynesty (or, equally \dynesty overestimates it). However, the relative difference shows that the differences are small compared to the log-evidence with a largest relative difference of less than 0.015\%. \Cref{fig:bbh_log_evidence} also shows the absolute difference in estimated uncertainty on the log-evidence from each sampler. The uncertainty from \pocomc is almost always larger than that from \dynesty (the difference is negative) which is consistent with previous studies that showed that \gls{ns} typically underestimates the uncertainty~\citep{2018BayAn..13..873H}.

\begin{figure}
    \centering
    \includegraphics{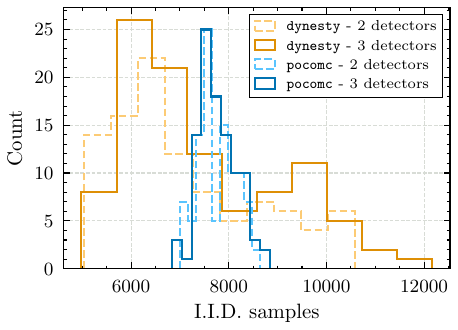}
    \caption{Number of \gls{iid}  posterior samples produce by \pocomc and \dynesty when analysing 100 simulated \gls{bbh} events. \pocomc was configured to return weighted samples with an \gls{ess} of 10,000, rejection sampling was then used to obtain \gls{iid} samples.}
    \label{fig:bbh_n_samples}
\end{figure}

\begin{figure}
    \centering
    \includegraphics{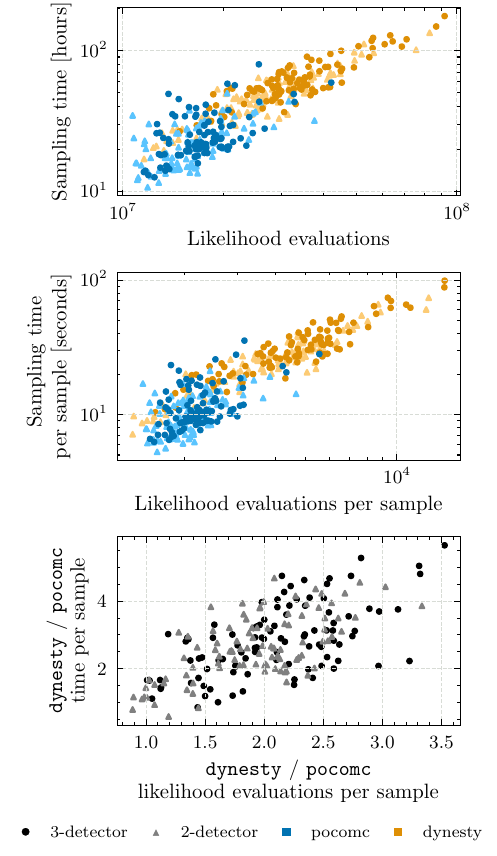}
    \caption{Comparison of the number of likelihood evaluations required and total wall time when sampling using \pocomc (blue) and \dynesty (orange) in two- and three-detector configurations. Circular markers denote three-detector results and triangular markers denote two-detector results. All analyses were run using four CPU cores. The average efficiency improvement for two-detectors (three-detectors) is \efficiencyimprovementtwodet (\efficiencyimprovementthreedet) or, in terms of wall time \timeimprovementtwodet (\timeimprovementthreedet)}
    \label{fig:bbh_comparison_dynesty}
\end{figure}

In \cref{fig:bbh_n_samples,fig:bbh_comparison_dynesty}, we compare the efficiency and cost of the different analyses. For both samplers, we track the total number of likelihood evaluations and total wall time for each analysis. We note however that the wall time is less reliable as the analyses may have run on different hardware.
Additionally, since the analyses do not produce the same number of samples (see \cref{fig:bbh_n_samples}), we compute the per-sample statistics.
We present these results in \cref{fig:bbh_comparison_dynesty} for both the two- and three-detector runs.
This shows that with chosen settings \pocomc is between two to seven times more efficient in terms of likelihood evaluations than \dynesty, which translates to a similar reduction in total wall time.
The average efficiency improvement for two-detectors (three-detectors) is \efficiencyimprovementtwodet (\efficiencyimprovementthreedet) or, in terms of wall time \timeimprovementtwodet (\timeimprovementthreedet) per sample.

\subsection{Binary neutron star analyses}\label{sec:bns}

We consider two regimes for \gls{bns} inference: fast inference suitable for low-latency analyses and slower, more robust inference for publication-quality analyses. In both cases, we use the \texttt{IMRPhenomPv2}~\citep{Hannam:2013oca} waveform approximant (optionally with tidal parameters~\citep{Dietrich:2019kaq}) and leverage reduced-order modelling with \gls{roq} to accelerate the gravitational-wave likelihood~\citep{Smith:2016qas}. Specifically, we use the \gls{roq} bases described in \cite{Morisaki:2023kuq} which available are in \cite{morisaki_2024_14279382} and are compatible with \bilby.

We consider the same two- and three-detector configurations described previously, and simulate a 128-second GW190425-like \gls{bns} signal using \texttt{IMRPhenomPv2\_NRTidalv2}. The signal is simulated with precessing spins and non-zero tidal parameters, the full set of injection parameters is described in \cref{tab:bns_parameters}.

For the low-latency regime, we use more aggressive settings, excluding tidal parameters and perform analyses with precessing or aligned spins. This reduces the number of binary parameters from 17 to 15 when considering precession, and to 11 when assuming aligned spins. Further speed-ups could be achieved by using techniques such as those described in \cite{Morisaki:2020oqk} but since this would apply to both samplers, we do not include them here. We also use less stringent sampler settings for aligned analyses with \pocomc, see \cref{tab:settings_dynesty_bbh}, and similarly to the \gls{bbh} analyses, we also analyse the same signal using \dynesty. We enable phase and distance marginalization for all analyses. We also restrict the prior on polarization angle to $[0, \pi/2)$ since the \texttt{IMRPhenomP} family of waveforms does not include higher-order modes and consider spin magnitudes up to 0.4~\citep{Harry:2018hke}.

\begin{figure}
    \centering
    \includegraphics[width=0.8\linewidth]{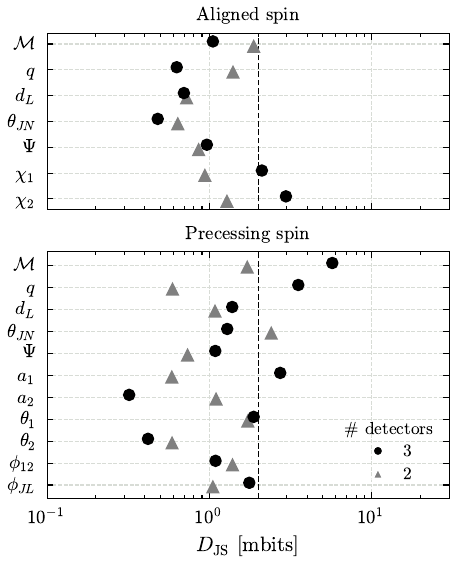}
    \caption{Mean \gls{jsd} in millibits between posterior distributions obtained with \dynesty and \pocomc when analysing a simulated \gls{bns} signal as described in \cref{sec:bns}. Results are shown for analyses with aligned and precessing spins, in both cases tidal effects are ignored, and for two- and three-detectors networks. The \gls{jsd} is computed 10 times per analysis using 5000 samples and the error bars show the standard deviation. The parameters are defined in \cref{tab:priors}.  The vertical dashed line shows the threshold proposed in \protect\cite{Ashton:2021anp}}
    \label{fig:bns_results}
\end{figure}

We compute the \gls{jsd} between the posterior samples produced by each sampler, comparing only cases where the priors and parameters being sampled are the same. \Cref{fig:bns_results} shows the results without tidal parameters and those with tidal parameters are shown in \cref{app:additional_results}.
These results show that, unlike for the \gls{bbh} analyses, the \gls{jsd} values are broadly consistent for both samplers irrespective of the number of detectors and choice of aligned or precessing spins.
However, when including tidal parameters in the analyses we observe larger differences and the inferred mass posteriors differ more than the analyses without tidal parameters.
We cannot rule out the possibility that the \dynesty results are not well converged but this suggests further investigation would be need if \pocomc were to be used for publication-quality \gls{bns} analyses.

We compare the sampling efficiency and cost, measured in terms of total number of likelihood evaluations and wall time, for each sampler and the various configurations for two-detector network in \cref{tab:bns_stats_2det}.
As expected, the results show a clear correlation between the sampling cost and complexity of the physics being modelled. However, there are significant differences between the two samplers; \pocomc is two to three times more efficient than \dynesty for the precessing analyses and the full precessing analyses with tidal parameters using \pocomc are comparable in cost to the aligned-spin analyses without tides using \dynesty.
This efficiency improvement could allow low-latency analyses to included additional physics without increasing their cost compared to using \dynesty.


\begin{table*}
\centering
\caption{Comparison of the total number of likelihood evaluations, wall times and likelihood evaluation per sample when analysing a GW190425-like \gls{bns} using \pocomc and \dynesty with two-detector network. All analyses were run using 16 CPU cores and results are averaged over two runs. The equivalent table for a three-detector network is included in \cref{app:additional_results}.}
\label{tab:bns_stats_2det}
\small

    \begin{tabular}{lcccccc}
    \toprule
    Sampler & Precession & Tides & Likelihood evaluations & Wall time [min] & Likelihood evaluations per sample \\
    \midrule
    \rowcolor{lightgrey}\texttt{dynesty} & \xmark & \xmark & $3.3 \times 10^{7}$ & 167.3 & 7273 \\
\texttt{pocomc} & \xmark & \xmark & $2.4 \times 10^{6}$ & 34.99 & 589.7 \\
\rowcolor{lightgrey}\texttt{dynesty} & \xmark & \cmark & $3.8 \times 10^{7}$ & 295.9 & 6686 \\
\texttt{pocomc} & \xmark & \cmark & $4.5 \times 10^{6}$ & 34.21 & 1140 \\
\rowcolor{lightgrey}\texttt{dynesty} & \cmark & \xmark & $4.1 \times 10^{7}$ & 303 & 7174 \\
\texttt{pocomc} & \cmark & \xmark & $1.9 \times 10^{7}$ & 272.8 & 2417 \\
\rowcolor{lightgrey}\texttt{dynesty} & \cmark & \cmark & $3.3 \times 10^{7}$ & 346.1 & 5395 \\
\texttt{pocomc} & \cmark & \cmark & $2.0 \times 10^{7}$ & 153.4 & 2556 \\
    \bottomrule
    \end{tabular}
    
\end{table*}

\section{Application to real data}\label{sec:real_data}

To further demonstrate the robustness of \pocomc we analyse two gravitational-wave events from the first three \gls{lvk} observing runs~\cite{LIGOScientific:2018mvr,LIGOScientific:2020ibl,LIGOScientific:2021usb,KAGRA:2021vkt}. We base our analyses on those performed in \glspl{gwtc}-2.1 and -3~\citep{LIGOScientific:2021usb,KAGRA:2021vkt} and use the strain data that is publicly available via GWOSC~\citep{LIGOScientific:2019lzm,KAGRA:2023pio}.
We compare our results to the samples included in the data releases for these catalogues~\citep{ligo_scientific_collaboration_and_virgo_2022_6513631,ligo_scientific_collaboration_and_virgo_2023_8177023}. All analyses are run with twice per event and using 16 cores.
Code to reproduce all of the analyses and the results are available in the accompanying data release~\citep{code_release,data_release}.

These analyses pose additional challenges compared to the simulated events, since the data may not be strictly stationary and Gaussian and additional calibration parameters described in \cref{sec:gwinference} must be included. We find that for these analyses \pocomc performs better without preconditioning using the normalizing flow---sampling is slow and prone to over-constraining the posterior distribution. We attribute this to the increased dimensionality of the problem, which makes effectively training the normalizing flow with the same number of samples challenging. We leave addressing this to future work.

\begin{figure}
    \centering
    \includegraphics[width=\linewidth]{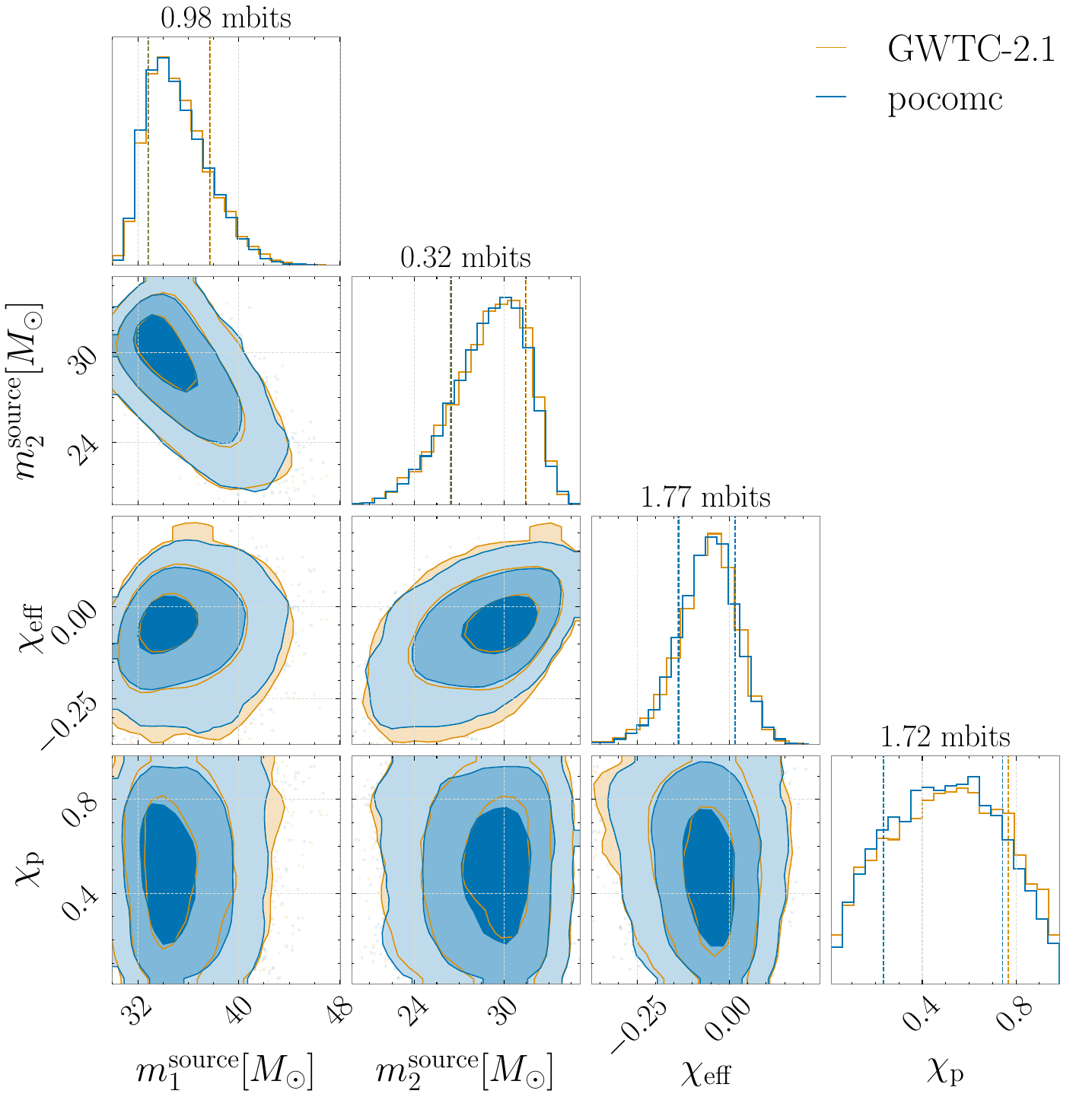}
    \caption{Inferred source frame component masses $m_1^\textrm{source}$ and $m_2^\textrm{source}$, effective spin $\chi_\textrm{eff}$ and precession parameter $\chi_\textrm{p}$ for GW150914\_095045 using \pocomc (blue). Results are compared against those released by the \gls{lvk} in \gls{gwtc}-2.1 (blue). The two-dimensional plots show the 1-, 2- and 3-$\sigma$ contours. The vertical dashed line shows the 1-$\sigma$ confidence interval. The \gls{jsd} in mbits between the 1--dimensional posterior distributions is quoted above each histogram.}
    \label{fig:GW150914_intrinsic}
\end{figure}

\begin{figure}
    \centering
    \includegraphics[width=\linewidth]{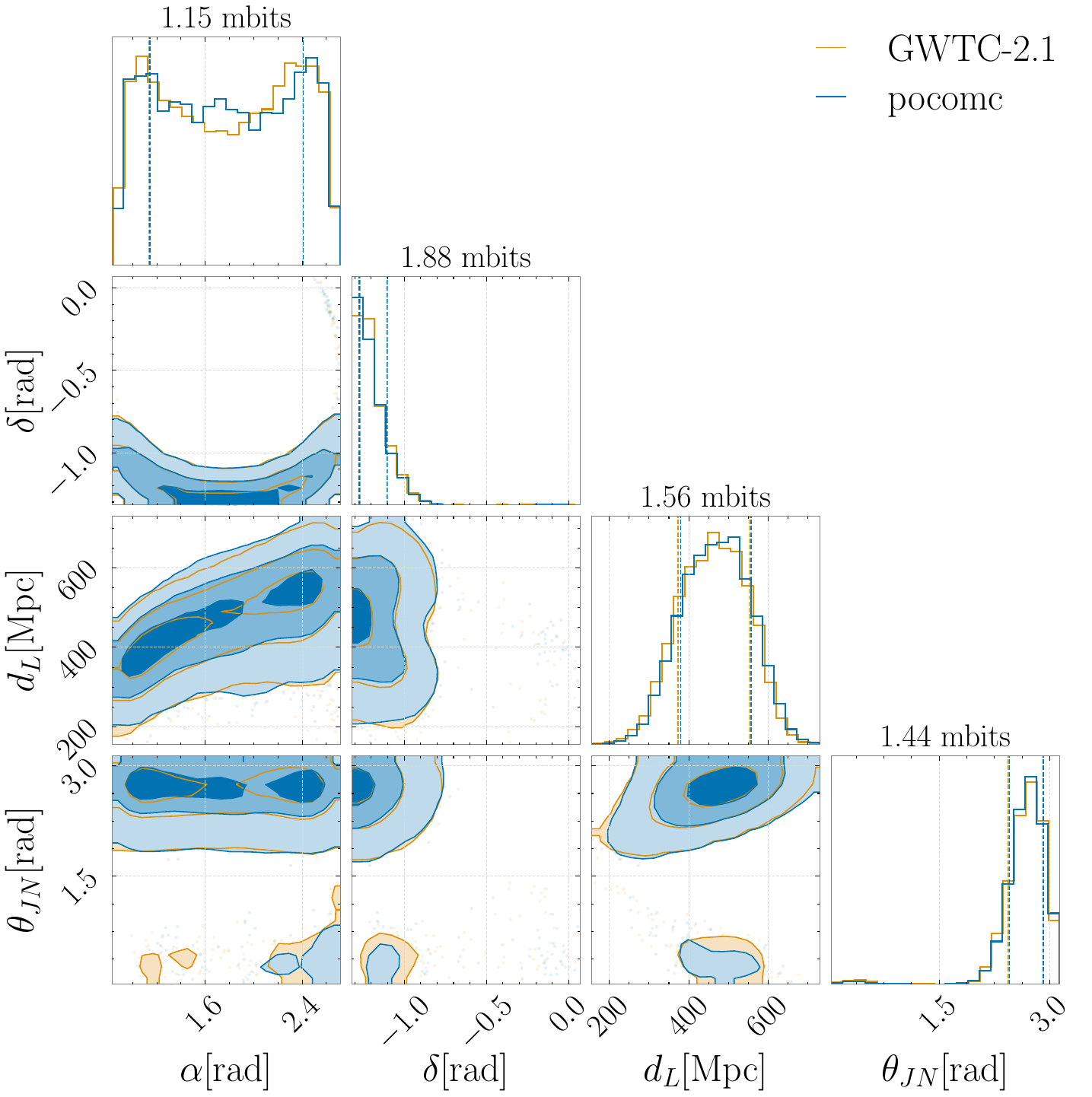}
    \caption{Inferred right ascension $\alpha$, declination $\delta$, luminosity distance $d_\textrm{L}$ and inclination angle $\theta_\textrm{JN}$ for GW150914\_095045 using \pocomc (red). Results are compared against those released by the \gls{lvk} in \gls{gwtc}-2.1 (orange). The two-dimensional plots show the 1-, 2- and 3-$\sigma$ contours. The vertical dashed line shows the 1-$\sigma$ confidence interval. The \gls{jsd} in mbits between the 1--dimensional posterior distributions is quoted above each histogram.}
    \label{fig:GW150914_location}
\end{figure}

GW150914\_095045 was the first \gls{bbh} detected by the \gls{lvk} and is considered a `vanilla' \gls{bbh} with source frame component masses of $36\;\textrm{M}_\odot$ and $29\;\textrm{M}_\odot$~\cite{LIGOScientific:2016vbw}. With \pocomc, the analyses took on average \checkme{9} hours to complete, produced \checkme{7781} samples after rejection sampling, and required of order \checkme{20} million likelihood evaluations to converge.
The inferred masses and spin parameters and subset of the extrinsic parameters are shown in \cref{fig:GW150914_intrinsic,fig:GW150914_location}.
These show overall agreement between the results obtained with \pocomc and included in the \gls{gwtc}-2.1 data release.
This demonstrates that \pocomc is robust to the inclusion of additional calibration parameters.
Furthermore, comparing to the results shown in \cref{fig:bbh_comparison_dynesty}, we see that the inclusion of the calibration parameters has not significantly impacted the number of likelihood evaluations \pocomc requires.

\begin{figure}
    \centering
    \includegraphics[width=\linewidth]{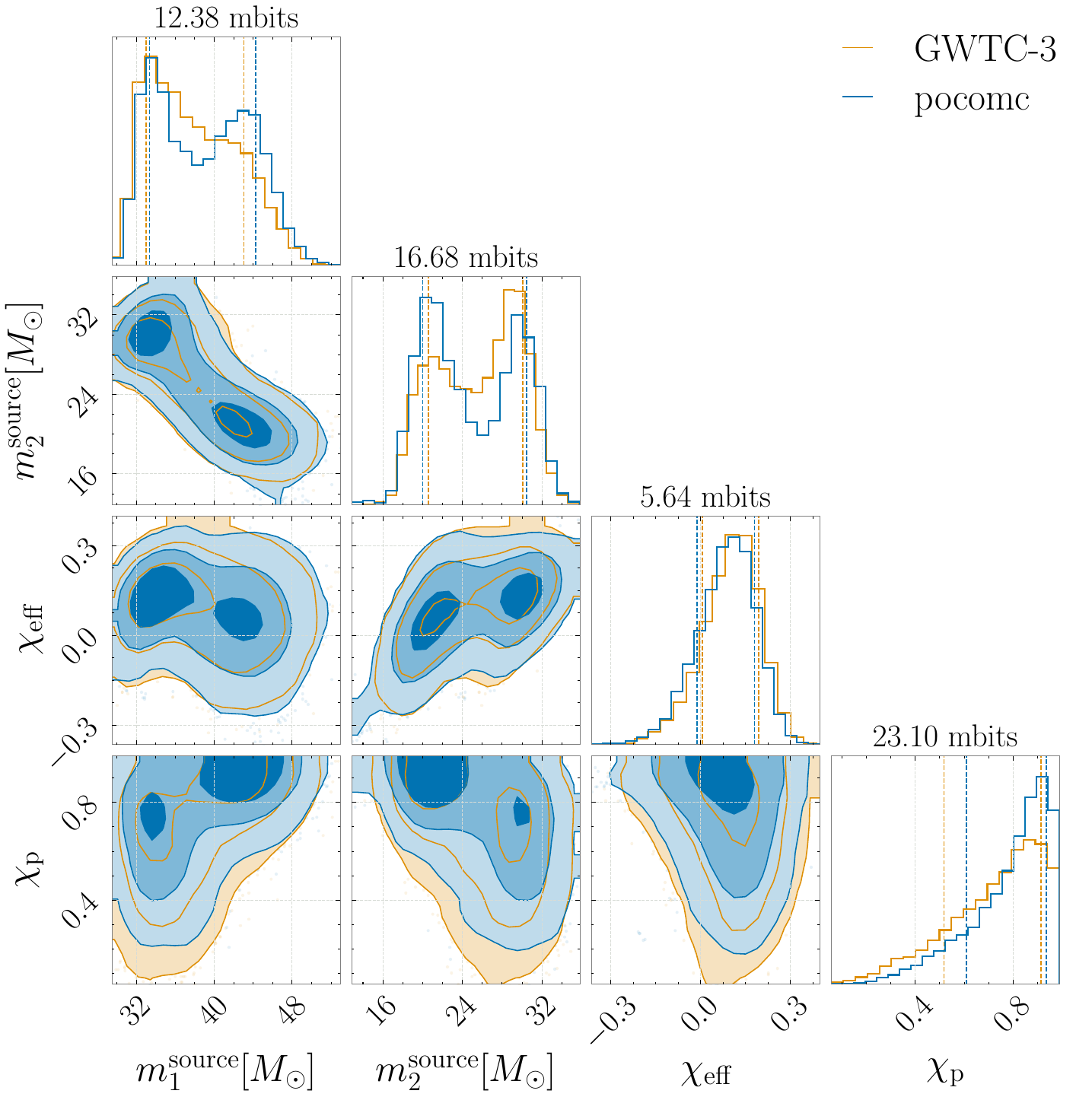}
    \caption{Inferred source frame component masses $m_1^\textrm{source}$ and $m_2^\textrm{source}$, effective spin $\chi_\textrm{eff}$ and precession parameter $\chi_\textrm{p}$ for GW200129\_065458 using \pocomc (blue). Results are compared against those released by the \gls{lvk} in \gls{gwtc}-3 (orange). The two-dimensional plots show the 1-, 2- and 3-$\sigma$ contours. The vertical dashed line shows the 1-$\sigma$ confidence interval. The \gls{jsd} in mbits between the 1--dimensional posterior distributions is quoted above each histogram.}
    \label{fig:GW200129_intrinsic}
\end{figure}

\begin{figure}
    \centering
    \includegraphics[width=\linewidth]{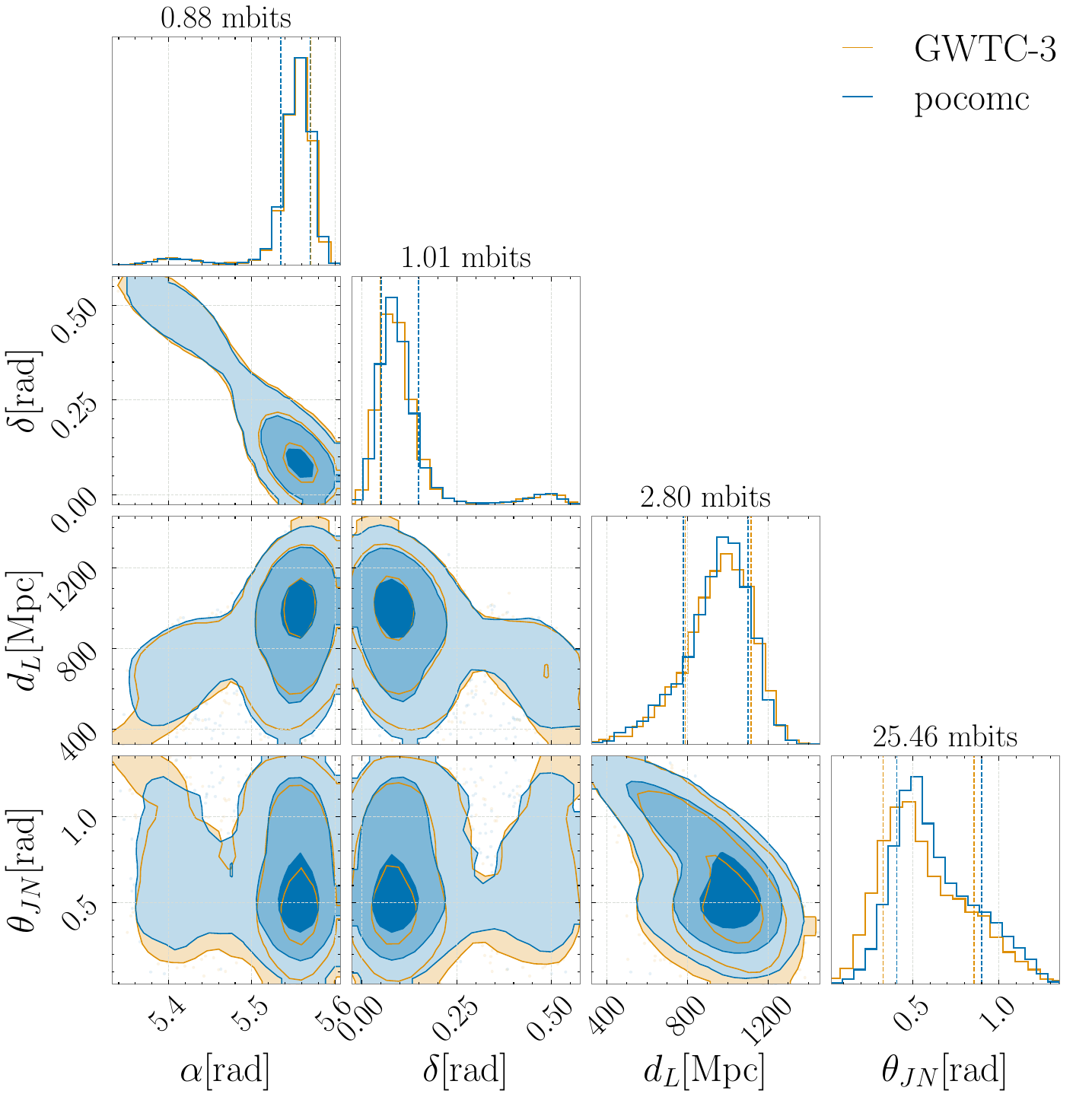}
    \caption{Inferred right ascension $\alpha$, declination $\delta$, luminosity distance $d_\textrm{L}$ and inclination angle $\theta_\textrm{JN}$ for GW200129\_065458 using \pocomc (red). Results are compared against those released by the \gls{lvk} in \gls{gwtc}-3 (blue). The two-dimensional plots show the 1-, 2- and 3-$\sigma$ contours. The vertical dashed line shows the 1-$\sigma$ confidence interval. The \gls{jsd} in mbits between the 1--dimensional posterior distributions is quoted above each histogram.}
    \label{fig:GW200129_location}
\end{figure}

GW200129\_065458 was a three-detector \gls{bbh} merger detected in the third \gls{lvk} observing run that shows evidence for spin-precession~\citep{Hannam:2021pit} and has been the subject of many analyses, e.g. \cite{Payne:2022spz,Macas:2023wiw}. When analysed with \pocomc, the analyses took on average \checkme{21} hours, required approximately \checkme{24 million} likelihood evaluations and produced \checkme{7277} samples after rejection sampling.
The posterior distributions from this analysis are compared to those included in the \gls{gwtc}-3 data release~\citep{ligo_scientific_collaboration_and_virgo_2023_8177023} in \cref{fig:GW200129_intrinsic,fig:GW200129_location}, our results are consistent with those produced by the \gls{lvk} but find additional support for the heavier mode at approximately $40\;\textrm{M}_{\odot}$ and for larger values of the precessing spin parameter, $\chi_\textrm{p}$.
The extrinsic parameters show better agreement.
Upon further examination, we find various differences in the analyses that could explain the differences in inferred intrinsic parameters. The Bayesian evidences for the noise-only hypothesis differ between the two results, suggesting differences in the data being analysed. Whilst both analyses use the same codes and waveform approximant, the exact versions differ, in particular the version of \texttt{IMRPhenomXPHM} used in our analyses contains several updates compared to \gls{gwtc}-3.

\section{Conclusions}\label{sec:conclusions}

In this work, we have demonstrated that \gls{smc} is a robust tool for performing gravitational-wave inference for \gls{cbc} signals and a viable alternative to \gls{ns}.
In particular, we used \pocomc, an \gls{smc} sampler that implements \gls{ps}, an extension to the standard \gls{smc} that addresses some of its main shortcomings, and applied it to a range of different \gls{cbc} analyses.

Initially, we test \pocomc on a series of simulated signals: 100 \gls{bbh} signals and one GW190425-like \gls{bns}. From these results, we conclude that:
\begin{itemize}
    \item{\pocomc can robustly perform inference for \gls{bbh} signals in two- and three-detector networks using waveforms that include precession and higher-order multipoles;}
    \item{compared to the standard sampler used by the \gls{lvk}, \dynesty, \pocomc, is on average \efficiencymean times more efficient and \timemean times faster whilst returning a more consistent number of samples;}
    \item{when applied to \gls{bns} signals, \pocomc is significantly more efficient than \dynesty and could therefore enable more complete analyses with the same latency.}
\end{itemize}

In addition to these simulated tests, we also reanalyse two real events from \glspl{gwtc}-2.1 and -3 with \pocomc: GW150914 and GW200129.
These analyses include additional parameters, 20 per-detector, to account for the inherent uncertainties in real data and these make inference more challenging.
We show that \pocomc is suitable for such analyses, producing results that are consistent with those released by the \gls{lvk}, but this required disabling the preconditioning with normalizing flows.
Further worked is need to improve the preconditioning to make it applicable to such analyses and to extensively verify that the settings used are robust across a wide range of events.

Looking forward, \gls{smc} should be considered when developing analyses for the next generation of ground- and space-based gravitational-wave detectors. In addition to this, \gls{smc} can easily leverage differentiable likelihoods via gradient-based Markov kernels such as Hamiltonian Monte Carlo~\citep{Neal:2011}, whilst also improving efficiency of such kernels for multimodal problems and allowing for adaptive tuning of the kernel parameters~\citep{buchholz2021adaptive}. This makes \gls{smc} a prime candidate for development of gradient-based analyses.

\section*{Acknowledgements}

The authors thank Ian Harry and John Veitch for providing feedback on the manuscript. MJW thanks Konstantin Leyde and Rahul Dhurkunde for helpful discussions, Charlie Hoy for advice on generating the \gls{bbh} injections and Patricia Schmidt for advice on using \texttt{IMRPhenomPv2\_NRTidalv2}.
MJW acknowledges support from ST/X002225/1, ST/Y004876/1 and the University of Portsmouth.
MN and US acknowledge funding from NSF Award Number 2311559, and from the U.S. Department of Energy, Office of Science, Office of Advanced Scientific Computing Research under Contract No. DE-AC02-05CH11231 at Lawrence Berkeley National Laboratory to enable research for Data-intensive Machine Learning and Analysis.

This document has been assigned LIGO Document No. LIGO-P2500231.

The authors are grateful for computational resources provided by the LIGO Laboratory and Cardiff University and supported by National Science Foundation Grants PHY-0757058 and PHY-0823459 and STFC grants ST/I006285/1 and ST/V005618/1. Additional numerical computations were carried out on the SCIAMA High Performance Compute (HPC) cluster which is supported by the ICG and the University of Portsmouth and on resources of the National Energy Research Scientific Computing Center (NERSC). This material is based upon work supported by NSF's LIGO Laboratory which is a major facility fully funded by the National Science Foundation.

This work made use of the following software packages: \texttt{bilby}~\citep{colm_talbot_2025_15059020}, \texttt{bilby\_pipe}~\citep{Ashton:2018jfp}, \texttt{corner}~\citep{corner}, \texttt{dynesty}~\citep{Speagle:2019ivv}, \texttt{LALSuite}~\citep{lalsuite,swiglal}, \texttt{matplotlib}~\citep{Hunter:2007}, \texttt{NumPy}~\citep{numpy}, \texttt{pandas}~\citep{reback2020pandas}, \texttt{PESummary}~\citep{Hoy:2020vys}, \texttt{pocomc}~\citep{karamanis2022pocomc}, \texttt{SciPy}~\citep{2020SciPy-NMeth}, \texttt{seaborn}~\citep{Waskom2021}.

\section*{Data Availability}

Interferometric data for the gravitational-wave events analysed and posterior samples produced by the LIGO-Virgo-KAGRA Collaboration are available via GWOSC~\citep{LIGOScientific:2019lzm,KAGRA:2023pio,ligo_scientific_collaboration_and_virgo_2022_6513631,ligo_scientific_collaboration_and_virgo_2023_8177023}.
The data release including code to reproduce the figure and analyses as well as the simulated data and the full results is available online \cite{code_release,data_release} and we also provide documentation for the data release at \url{http://gw-smc.michaeljwilliams.me}.
\pocomc is available at \url{https://github.com/minaskara/pocomc} and via PyPI.
The \pocomc-\bilby interface is available at \url{https://github.com/mj-will/pocomc-bilby}~\citep{pocomc_bilby} and via PyPI.


\bibliographystyle{mnras}
\bibliography{main}



\appendix

\section{Simulation details and sampler settings}\label{app:priors}

In this section, we provide additional details about the simulated data and sampler settings used throughout this work. Any settings not specified here can be found in the accompanying data release~\citep{}.

\Cref{tab:priors} details the priors used in \cref{sec:bbh} when analysing the \gls{bbh} signals in simulated data. The priors are based on those used in \cite{Williams:2021qyt}~\cite{Williams:2021qyt}. Injections were made using a custom script to ensure data consistency between the different analyses, the script is available in the data release. All signals simulated using \texttt{IMRPhenomXPHM}~\citep{Pratten:2020ceb} and injected into \checkme{512} seconds of Gaussian noise coloured according to the \gls{psd} of the desired detector. Analyses were performed using an 8-second window with two seconds on data after the trigger time.


\begin{table}
    \caption{Prior distributions used for inference in \cref{sec:bbh}. For definitions of each parameter, see Appendix E of \protect\cite{Romero-Shaw:2020owr}. Table reproduced from \protect\cite{Williams:2023iss}.}
    \label{tab:priors}
    \begin{tabular}{|l|l|c|} \hline
        \centering
    Parameters &  Prior & Bounds \\ \hline
    $\mathcal{M}$ & Uniform & $[25, 35]\;\textrm{M}_\odot$ \\
    $q$ & Uniform & $[0.125, 1.0]$ \\
    $d_{\text{L}}$ & Uniform in co-moving volume & $[100, 2000]\;\textrm{Mpc}$ \\
    $\alpha$ & Uniform & $[0, 2 \pi]$ \\
    $\delta$ & Cosine & - \\
    $t_{\text{c}}$ & Uniform around trigger time & $[-0.1, 0.1]$ \\
    $\theta_{\text{JN}}$ & Sine & - \\
    $\psi$ & Uniform & $[0, \pi]$ \\
     $\phi_{c}$ & Uniform & $[0, 2 \pi]$ \\
    $a_{i}$ & Uniform & $[0, 0.99]$ \\
    $\theta_i$ & Sine & - \\
    $\Delta\phi$ & Uniform & $[0, 2 \pi]$ \\
    $\phi_{\text{JL}}$ & Uniform & $[0, 2 \pi]$ \\
        \hline
    \end{tabular}
\end{table}

\begin{table}
    \caption{Parameters for \gls{bns} injection generated using \texttt{IMRPhenomPv2\_NRTidalv2}. The chirp mass is defined in detector frame. For definitions of each parameter, see Appendix E of \protect\cite{Romero-Shaw:2020owr}.}
    \centering
    \begin{tabular}{|l|c|}
\hline
Parameters & Value \\
\hline
$\mathcal{M} [M_{\odot}]$ & 1.45 \\
$q$ & 0.93 \\
$a_{1}$ & 0.01 \\
$a_{2}$ & 0.01 \\
$\theta_{1} [\mathrm{rad}]$ & 0.28 \\
$\theta_{2} [\mathrm{rad}]$ & 0.30 \\
$\phi_{12} [\mathrm{rad}]$ & 0.48 \\
$\phi_{JL} [\mathrm{rad}]$ & 0.33 \\
$d_{L} [\mathrm{Mpc}]$ & 45.00 \\
$\delta [\mathrm{rad}]$ & 0.31 \\
$\alpha [\mathrm{rad}]$ & 4.70 \\
$\theta_{JN} [\mathrm{rad}]$ & 2.75 \\
$\Psi [\mathrm{rad}]$ & 0.03 \\
$\phi [\mathrm{rad}]$ & 0.67 \\
$\lambda_{1}$ & 149.35 \\
$\lambda_{2}$ & 984.62 \\
$t_{c} [\mathrm{s}]$ & 0.06 \\
\hline
\end{tabular}

    \label{tab:bns_parameters}
\end{table}

\Cref{tab:bns_parameters} list the parameters used for simulating the \gls{bns} signal used in \cref{sec:bns}. The signal was simulated using \texttt{IMRPhenomPv2\_NRTidalv2}~\citep{Hannam:2013oca,Dietrich:2019kaq} and the functionality in \texttt{bilby\_pipe} and injected into 128 seconds of coloured Gaussian noise. With these parameters and assuming O3 sensitivity, the matched filter \gls{snr} in each detector: \checkme{12.04} in H1, \checkme{14.14} in L1 and \checkme{8.56} in V1 and the overall network \gls{snr} is \checkme{20.45}

The settings used for all analysis with \dynesty are detailed in \cref{tab:settings_dynesty_bbh}, these settings have been used extensively in \gls{lvk} analyses; see e.g. \cite{LIGOScientific:2024elc}. However, unlike the \gls{lvk} analyses that use two or more sampling runs per analysis, we only use a single sampling run per injection. This reduces the computational cost but makes the results less robust to statistical variations from sampling.

\begin{table}
    \caption{\pocomc settings for various analyses described in \cref{sec:simlated_data}. For details about the different settings see the \pocomc documentation.}
    \label{tab:settings_pocomc_bbh}
    \centering
    \begin{tabular}{|c|c|c|c|}
         \hline
         Setting & \gls{bbh} & \gls{bns} -- fast & \gls{bns} -- slow\\ \hline
         \texttt{n\_effective} & 2000  & 1000 & 2000\\
         \texttt{n\_active} & 1000 & 500 & 1000\\
         \texttt{n\_steps} & 100 & 50 & 50 \\
         \texttt{n\_max\_steps} & 1000 & 500 & 500 \\ 
         \texttt{n\_samples} & 10,000 & 5000 & 10,000 \\
         \hline
    \end{tabular}

\end{table}

\begin{table}
    \caption{\dynesty settings for all analyses. For details about the different settings see \dynesty and \bilby documentation.}
    \label{tab:settings_dynesty_bbh}
    \centering
    \begin{tabular}{|c|c|}
         \hline
         Setting & Value \\ \hline
         \texttt{nlive} & 1000  \\
         \texttt{naccept} & 60 \\
         \texttt{sample} & \texttt{acceptance-walk} \\ \hline
    \end{tabular}
   
\end{table}

\FloatBarrier
\section{Additional results}\label{app:additional_results}

\begin{figure}
    \centering
    \includegraphics{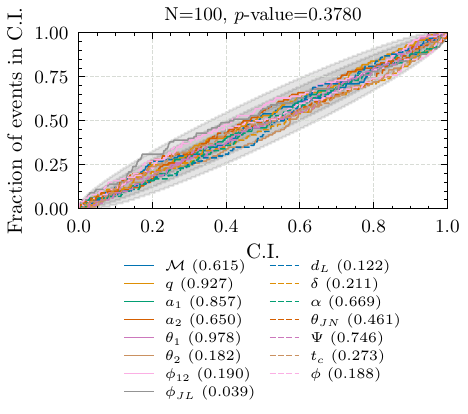}
    \caption{Probability-probability plot obtained using \pocomc to analyse 100 simulated \gls{bbh} signals in three-detector network consisting of LIGO Hanford, LIGO Livingston and Virgo. The shaded regions denote the 1-, 2- and 3-$\sigma$ bounds respectively. Per-parameter $p$-values are quoted, as well as the combined $p$-value computed using Fisher's method~\citep{fisher1970statistical}.}
    \label{fig:bbh:pp_test_3det}
\end{figure}

\begin{figure}
    \centering
    \includegraphics{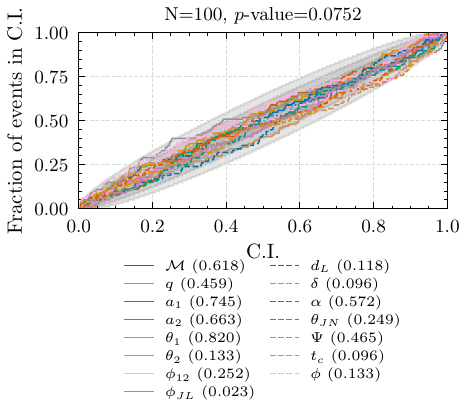}
    \caption{Probability-probability plot obtained using \dynesty to analyse 100 simulated \gls{bbh} signals in three-detector network consisting of LIGO Hanford, LIGO Livingston and Virgo. The shaded regions denote the 1-, 2- and 3-$\sigma$ bounds respectively. Per-parameter $p$-values are quoted, as well as the combined $p$-value computed using Fisher's method~\citep{fisher1970statistical}.}
    \label{fig:bbh:pp_test_3det_dynesty}
\end{figure}

\begin{figure}
    \centering
    \includegraphics{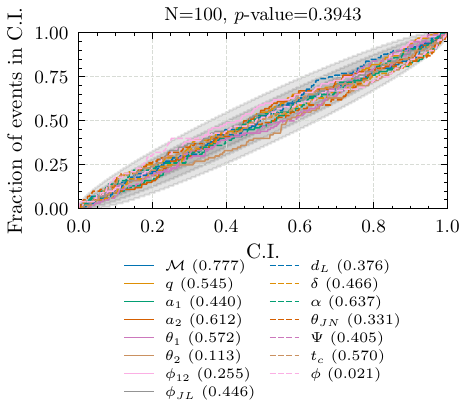}
    \caption{Probability-probability plot obtained using \dynesty to analyse 100 simulated \gls{bbh} signals in two-detector network consisting of LIGO Hanford and LIGO Livingston. The shaded regions denote the 1-, 2- and 3-$\sigma$ bounds respectively. Per-parameter $p$-values are quoted as well as the combined $p$-value computed using Fisher's method~\citep{fisher1970statistical}.}
    \label{fig:bbh:pp_test_2det_dynesty}
\end{figure}

In this section, we include additional results from the \glspl{pptest} described in \cref{sec:bbh}. \Cref{fig:bbh:pp_test_3det} shows the results of three-detector \gls{pptest} with \pocomc and \cref{fig:bbh:pp_test_2det_dynesty,fig:bbh:pp_test_3det_dynesty} show the results for the two- and three-detectors tests for \dynesty.
In all cases, the $p$-value is above 0.05, indicating the samplers pass the test.

\begin{figure}
    \centering
    \includegraphics[width=0.8\linewidth]{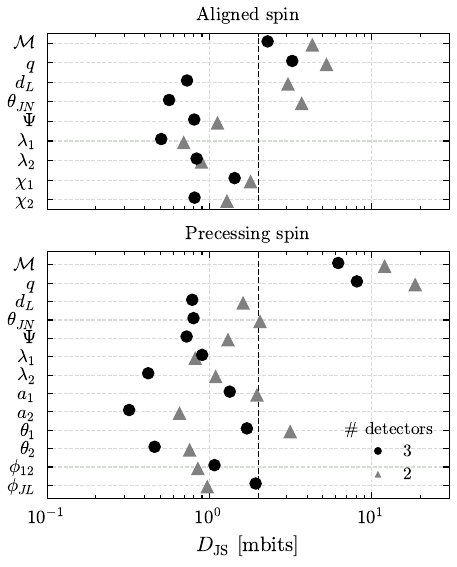}
    \caption{Mean \gls{jsd} in millibits between posterior distributions obtained with \dynesty and \pocomc when analysing a simulated \gls{bns} signal as described in \cref{sec:bns}. Results are shown for analyses with aligned and precessing spins, tidal effects, and for two- and three-detectors networks. The \gls{jsd} is computed 10 times per analysis using 5000 samples and the error bars show the standard deviation. The parameters are defined in \cref{tab:priors}.  The vertical dashed line shows the threshold proposed in \protect\cite{Ashton:2021anp}}
    \label{fig:bns_results_w_tides}
\end{figure}

\begin{table*}
\centering
\caption{Comparison of the total number of likelihood evaluations, wall times and likelihood evaluation per sample when analysing a GW190425-like \gls{bns} using \pocomc and \dynesty with two-detector network. All analyses were run using 16 CPU cores and results are averaged over two runs. The equivalent results for a two-detector network are shown \cref{tab:bns_stats_2det}.}
\label{tab:bns_stats_3det}
\small

    \begin{tabular}{lcccccc}
    \toprule
    Sampler & Precession & Tides & Likelihood evaluations & Wall time [min] & Likelihood evaluations per sample \\
    \midrule
    \rowcolor{lightgrey}\texttt{dynesty} & \xmark & \xmark & $3.7 \times 10^{7}$ & 199.6 & 7530 \\
\texttt{pocomc} & \xmark & \xmark & $3.1 \times 10^{6}$ & 23.72 & 748.6 \\
\rowcolor{lightgrey}\texttt{dynesty} & \xmark & \cmark & $3.0 \times 10^{7}$ & 181.4 & 5479 \\
\texttt{pocomc} & \xmark & \cmark & $4.2 \times 10^{6}$ & 35.78 & 1002 \\
\rowcolor{lightgrey}\texttt{dynesty} & \cmark & \xmark & $2.9 \times 10^{7}$ & 129.7 & 5942 \\
\texttt{pocomc} & \cmark & \xmark & $6.3 \times 10^{6}$ & 43.75 & 1895 \\
\rowcolor{lightgrey}\texttt{dynesty} & \cmark & \cmark & $2.5 \times 10^{7}$ & 169.9 & 4660 \\
\texttt{pocomc} & \cmark & \cmark & $1.8 \times 10^{7}$ & 143.6 & 2389 \\
    \bottomrule
    \end{tabular}
    
\end{table*}


\bsp	
\label{lastpage}
\end{document}